\documentclass[twocolumn,showpacs,preprintnumbers,amsmath,amssymb]{revtex4}

\usepackage{graphicx}
\usepackage{dcolumn}
\usepackage{bm}

\usepackage{graphicx}
\begin{document}

\preprint{APS/123-QED}

\title{Specific heat of segmented Heisenberg quantum spin chains in 
(Yb$_{1-x}$Lu$_x$)$_4$As$_3$}

\author{R. Matysiak}
\affiliation{Institute of Engineering and Computer Education,
University of Zielona G\'{o}ra, ul. prof. Z. Szafrana 4,
65-516 Zielona G\'{o}ra Poland}
\email{r.matysiak@eti.uz.zgora.pl}

\author{P. Gegenwart}
\affiliation{Max Planck Institute for Chemical Physics of Solids, 01187 Dresden, Germany}
\affiliation{I. Physikalisches Institut, Georg-August-Universit\"{a}t
G\"{oe}ttingen, Friedrich-Hund-Platz 1, 37077  G\"{oe}ttingen, Germany}

\author{A. Ochiai}
\affiliation{Center for Low Temperature Science, Tohoku University, 
Sendai 980-8578, Japan}

\author{M. Antkowiak, G. Kamieniarz}
\affiliation{Computational Physics Division, Faculty of Physics,
A. Mickiewicz University, ul. Umultowska 85, 61-614 Pozna\'n, Poland}
\email{gjk@amu.edu.pl} 

\author{F. Steglich}
\affiliation{Max Planck Institute for Chemical Physics of Solids, 01187 Dresden, Germany}

\date{\today}

\begin{abstract}
We report low-temperature specific heat, $C(T)$, measurements 
on (Yb$_{1-x}$Lu$_x$)$_4$As$_3$ with $x=0.01$ and $x=0.03$, where nonmagnetic Lu atoms 
are randomly distributed on antiferromagnetic $S=1/2$ Heisenberg chains with 
$J/k_{\mathrm B}=28$~K. The observed reduction of $C$ below 15 K with increasing 
$x$ is accurately described by quantum transfer matrix simulations without any 
adjustable parameter, implying that the system is an
excellent experimental realization of segmented quantum spin chains. 
Finite-size effects consistent with conformal-field theory predictions are 
leading to the formation of an effective low-energy gap. The size of the gap 
increases with Lu content and accounts for the impurity driven reduction of the specific heat. 
For both concentrations our results verify experimentally the low temperature scaling 
behavior established theoretically 
and also confirm the value of $J$ determined from pure Yb$_4$As$_3$.
\end{abstract}


\maketitle

\section{Introduction}

Antiferromagnetic (AF) spin chains have attracted a lot of theoretical and 
experimental interest due to their intrinsic quantum properties. 
Theoretical description of such simple systems is less complex compared to 
higher dimensional spin interactions and still their analysis is very useful. 
The ground state of integer spin chains was predicted 
disordered with a gap in the excitation spectrum \cite{Haldane}. 
The presence of the energy gap between the ground state and the 
lowest-excited states has been realized in real systems 
\cite{Renard,Walcerz} and confirmed in simulations 
\cite{Walcerz,RM-PRB97}. 

The importance of computer simulations in studies of the
low-dimensional quantum spin systems \cite{Boner,Delica} has recently 
increased as a result of progress in the methods of simulations and 
increasing computer power \cite{MB,JP_CM2001}.
For the spin chains the simulation techniques have 
brought reliable results of high accuracy, which can be 
verified sometimes by the exact theoretical solutions and can 
be also used to verify the approximate theoretical results 
\cite{Jonston,Shibata}. Among the various methods of
computer simulations the quantum 
transfer matrix (QTM) technique plays an important role. 
Its accuracy and advantages were presented in a number of 
publications~\cite{CPC2002,CMS2003,PRB-MA}. 

The theory of the ideally uniform S = 1/2 
AF Heisenberg chain is well established \cite{Jonston}.
An energy gap may also exist in these non-integer spin 
systems as a consequence of the effect of a staggered magnetic field \cite{Affleck}. 
A typical system displaying such behaviour is Yb$_4$As$_3$~\cite{Schmidt}. 
At high temperatures, this system displays a cubic metallic mixed valence phase. 
Upon cooling to below 300 K, it undergoes a charge ordering transition, coupled 
to a structural distortion, which leads to the formation of domains. 
In this phase magnetic Yb$^{3+}$ ions form one-dimensional chains. 

Low-temperature properties of Yb$_4$As$_3$ are described 
by effective $S=1/2$ spin chains with AF interactions and staggered 
field \cite{Kohgi2, Kohgi4, Shibata}. Previously, we have used the 
QTM method and the Bethe ansatz solution~\cite{Jonston} to reproduce 
the low-temperature specific heat of Yb$_4$As$_3$ at zero field, as well 
as in the presence of an external field, which induces a gap in the 
low-energy excitations~\cite{prb2009rm, app2010}.

In this paper, we investigate the reduction of the specific heat and the 
low-energy excitations in Yb$_4$As$_3$
with random dilution of the magnetic moments. Since magnetic Yb$^{3+}$ is 
chemically identical to non-magnetic Lu$^{3+}$ and the charge ordering in 
(Yb$_{1-x}$Lu$_x$)$_4$As$_3$  is retained for $x\leq 0.06$~\cite{Aoki},  
the non-magnetic Lu is uniformly distributed, leading to a statistical 
segmentation of the spin chains. Theoretically, it is expected that the 
low-frequency spectral response of the $S=1/2$ AF Heisenberg chain becomes 
gaped by spin segmentation~\cite{Haas} and the static properties fulfill 
a scaling law though we are not aware of 
quantitative comparison with experiments.

Below we present experimental zero-field specific heat and simulation 
results on (Yb$_{1-x}$Lu$_x$)$_4$As$_3$ ($x=0$, $0.01$, $0.03$). 
Experimentally, we observe a strong reduction of the low-temperature
specific heat. Using the same exchange coupling $J/k_B=28$ K as  
for pure Yb$_4$As$_3$ \cite{prb2009rm} and assuming a random distribution 
of the non-magnetic impurities with concentration $x$ we calculate the 
average energy gap. 
The gap depends strongly on the size of the finite segments. Although 
for small concentrations (of the order of 1\%) the average gap is small, 
its impact on the specific heat is strong below 10 K. Our calculation 
quantitatively 
explains the reduction of the specific heat and the experimental results 
confirm the scaling behavior expected for segmented Heisenberg spin chains. 

\section{Experimental details}

\begin{figure}
\begin{center}
\includegraphics*[scale=0.44]{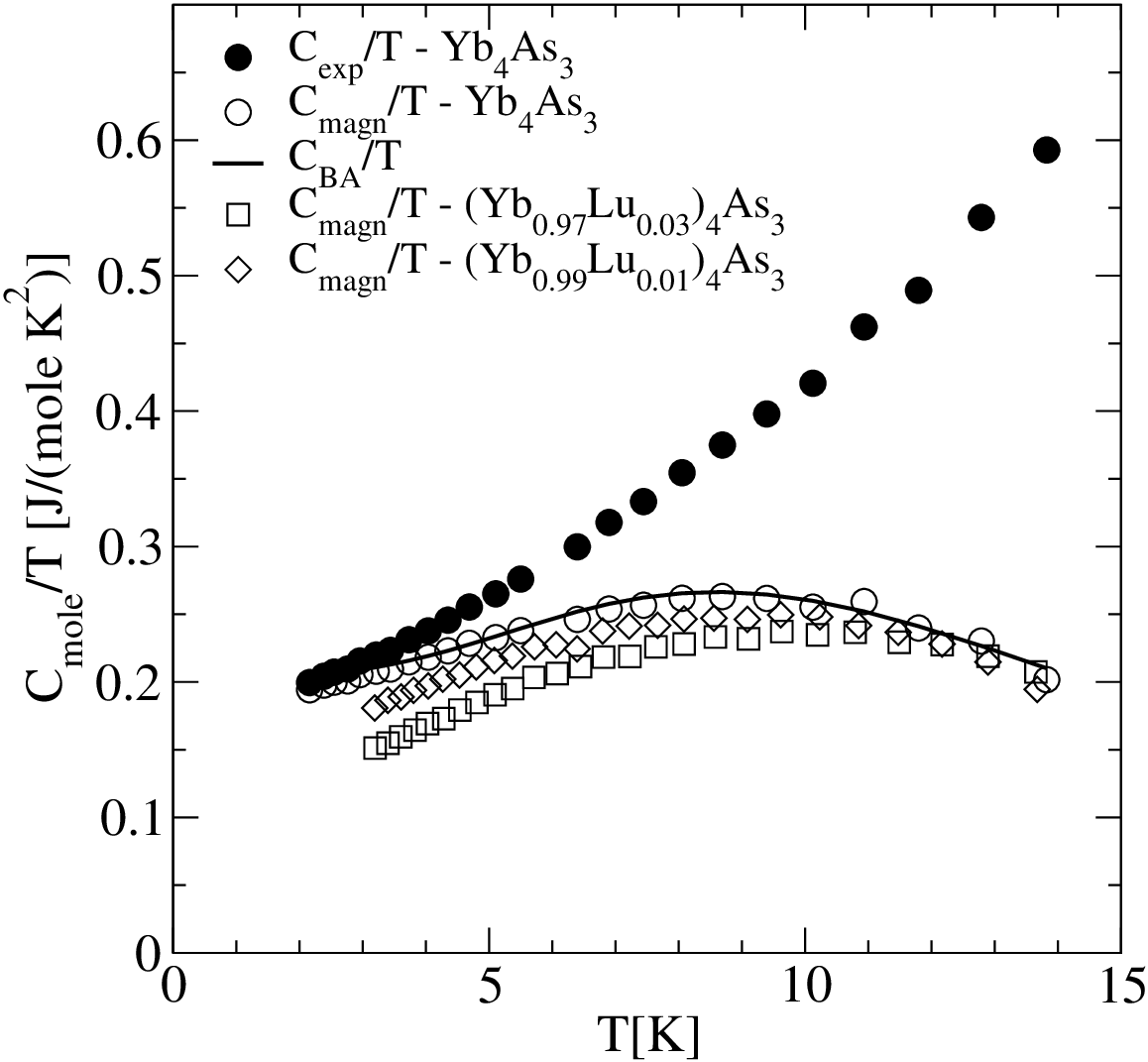}
\end{center}
\caption{Experimental results of zero-field specific heat
for pure Yb$_4$As$_3$ system and diluted (Yb$_{1-x}$Lu$_x$)$_4$As$_3$ 
for different impurity concentrations $x$.} 
\label{C_exp}
\end{figure}

The experiments have been performed on single crystals of 
(Yb$_{1-x}$Lu$_x$)$_4$As$_3$, characterized previously~\cite{Aoki,JJAP}. 
Small pieces of mass 9.6 mg ($x=0.01$) and 2.9 mg ($x=0.03$) have been 
investigated in a commercial micro-calorimeter from Oxford Instruments. 
In the previous paper \cite{prb2009rm} measurements of the specific heat 
were performed on the poly-domain sample of the undoped Yb$_4$As$_3$. 
The experimental results both for the pure and for the doped 
system are displayed in Fig. \ref{C_exp}. Experimental specific heat $C_{exp}$ consists 
of two components: the magnetic part $C_{magn}$ and its lattice counterpart 
$C_{ph}$. The magnetic contribution in zero field for Yb$_4$As$_3$ was 
determined theoretically within the Bethe ansatz (BA) solution expressed 
by the Pade approximants \cite{Jonston}.
From the comparison of the total specific heat with the Bethe ansatz 
estimation of the magnetic heat capacity, we have previously determined the 
phonon contribution to the heat capacity of 
Yb$_4$As$_3$ as 
$$C_{\rm ph}=\alpha T^3+\beta T^5$$ 
with 
$\alpha=1.11\times 10^{-3}$~J/(molK$^4$) and 
$\beta=4.9\times 10^{-6}$~J/(molK$^6$)~\cite{prb2009rm}. 

In Fig. \ref{C_exp} the BA results are plotted by continuous line and yield 
the optimal fit to the experimental part $C_{magn}$ represented by the 
open circles for the antiferromagnetic coupling $J/k_B=28$ K. In addition, 
the full circles show the total specific heat for the pure compound Yb$_4$As$_3$.

Since we do not expect a change of the phonon component by the small 
partial substitution of Yb by Lu, we have subtracted the same phonon contribution 
from the data in this study. Note, that we display the data per mol 
(Yb$_{1-x}$Lu$_x$)$_4$As$_3$, i.e. the heat capacity has not been rescaled 
to the amount of magnetic sites in the system. In Fig. \ref{C_exp} 
the diamonds and squares demonstrate the magnetic part of the specific heat 
measured for the system with impurities. The observed 
reduction of $C/T$ with $x$ is far larger than the natural reduction due to 
the $1$ or $3\%$ decrease of magnetic degrees of freedom for $x=0.01$ and 
$x=0.03$, respectively.

\section{Description of the model and the simulation technique}

Computer simulations of the zero-field specific heat of the 
diluted (Yb$_{1-x}$Lu$_x$)$_4$As$_3$ ($x=0.01$, $0.03$) systems
are based on the $S=1/2$ isotropic Heisenberg model for each 
segment consisting of finite number $N$ of spins:

\begin{equation}
 {\mathcal{H}} =  J \sum_{i=1}^{N}{\mathbf S}_{i} 
 {\mathbf S}_{i+1} .
\label{hamilt_iso}
\end{equation}
The thermodynamic properties of (Yb$_{1-x}$Lu$_x$)$_4$As$_3$ are 
described by a fixed value for the exchange coupling $J/k_B=28$ K which was 
found for the pure compound.
They are calculated from the derivatives of the free energy related 
to the partition function $\mathcal{Z}$, using the definition

\begin{equation}
\mathcal Z = {\mathbf{Tr}} \, e^{-\beta {\mathcal H}} .
\label{Z0}
\end{equation}

The partition function cannot
be calculated directly for large $N$ because of non-commuting operators 
in (\ref{hamilt_iso}). To eliminate this restriction, we look for 
systematic approximants to the partition function $\mathcal{Z}$, mapping 
the quantum system onto a classical one. 
We express Hamiltonian (\ref{hamilt_iso}) in terms
of the spin--pair operators ${\mathcal H}_{i,i+1}$ as a sum of two 
noncommuting parts \cite{Delica}

\begin{eqnarray} 
{\mathcal H} = {\mathcal H}^{odd} + {\mathcal H}^{even} =
\left( {\mathcal H}_{1,2} + \ldots + {\mathcal H}_{2i-1,2i} + \ldots \right)+
\nonumber \\
+\left( {\mathcal H}_{2,3} + \ldots + {\mathcal H}_{2i,2i+1} + \ldots \right) \, ,
\label{Hamiltonian3}
\end{eqnarray}
where each part is defined by the commuting components
${\mathcal H}_{i,i+1}$. The series of the classical approximants
can be found, using the general
Suzuki--Trotter formula \cite{Delica}. The partition function is 
calculated as the limit of the corresponding approximants

\begin{eqnarray}
\mathcal Z = \lim_{m \rightarrow \infty} \mathcal Z_m =
\nonumber \\
=\lim_{m \rightarrow \infty} \mathbf{Tr} \left[ \prod_{i=1}^{N/2}
{\mathcal V}_{2i-1,2i} \prod_{i=1}^{N/2} {\mathcal V}_{2i,2i+1}
\right]^m \, ,
\label{Z}
\end{eqnarray}
where ${\mathcal V}_{i,i+1} = e ^{-\beta {\mathcal H}_{i,i+1}/m}$,
$i=1,2, \cdots ,N$ and $m$ is a natural number (reffered to as the 
Trotter number). 

The approximant $\mathcal Z_m$ can be calculated numerically, 
without any restrictions on the value of $N$, by the quantum 
transfer--matrix (QTM) method. The computation of $\mathcal Z_m$ 
is possible for relatively small values of $m$, because of computer 
storage limitation, but for $m$ large enough the leading errors the finite-$m$ 
approximants are of the order of $1/m^2$ and therefore, extrapolations 
to $m \rightarrow \infty$ can be performed and the accurate estimates of 
$\mathcal Z$ can be obtained. 

To describe properties of the diluted system, we have to calculate 
the partition function and specific heat for the finite segments of the size $N$.  
We need to define the two vectors which act in the Hilbert space 
${\mathcal H}^{2m}$ \cite{Delica}:

\begin{equation}
\mid a \rangle = \sum_{\{ S^z \}} \prod_{r=1}^{2m} 
\delta_{S_{2r-1}^z , S_{2r}^z} \mid S_1^z \ldots S_{2m}^z \rangle
\end{equation}
 
\begin{equation}
\mid b \rangle = \sum_{\{ S^z \}} \prod_{r=1}^{2m} 
\delta_{S_{2r}^z , S_{2r+1}^z} \mid S_1^z \ldots S_{2m}^z \rangle
\end{equation}

\begin{figure}
\begin{center}
\includegraphics*[scale=0.44]{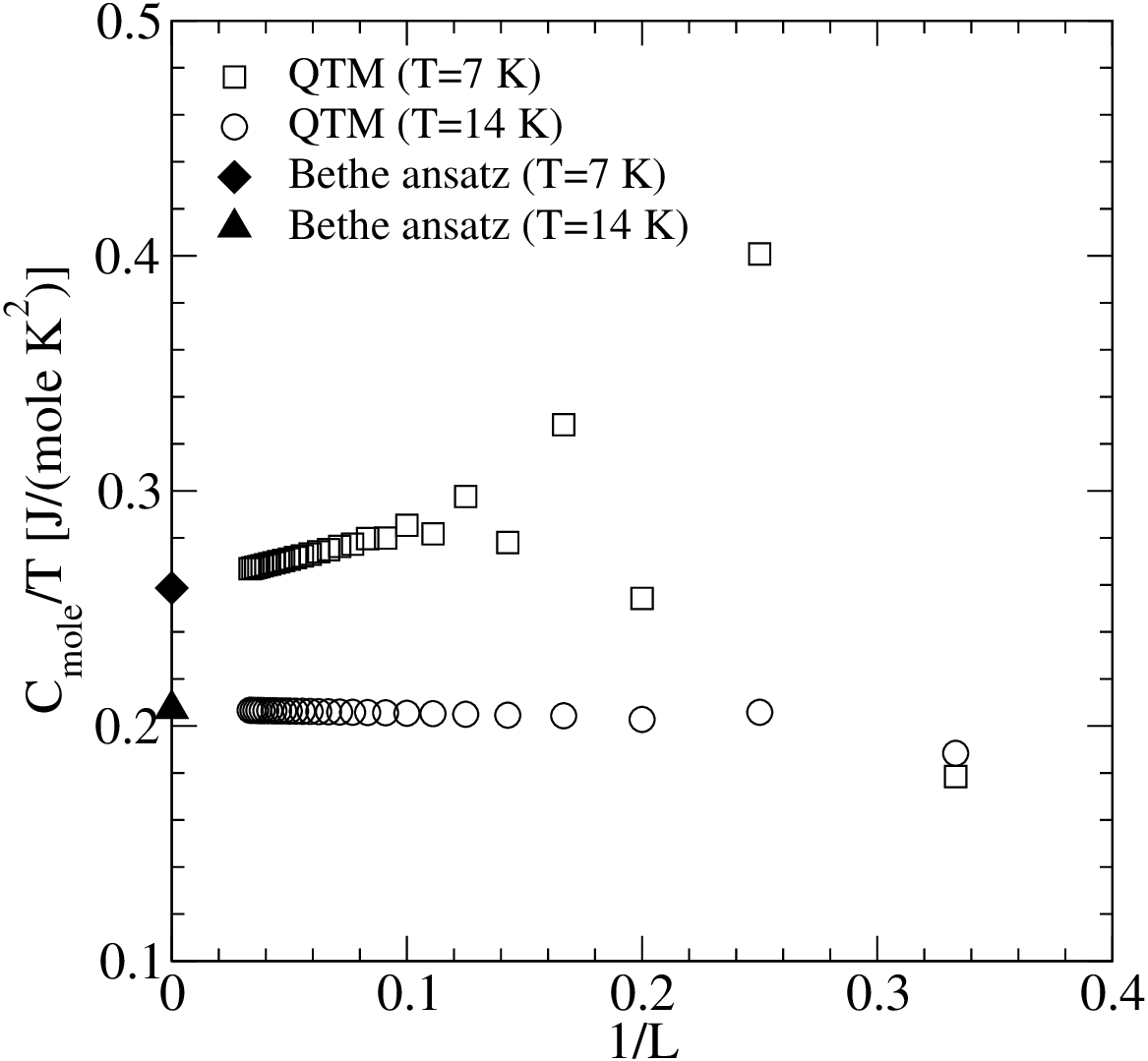}
\end{center}
\caption{Size dependence of zero-field specific heat calculated for 
finite segments with different number of sites $L$. The QTM data 
are shown by open symbols. The filled symbols represent the Bethe ansatz 
results 
corresponding to the macroscopic limit.}
\label{c_l}
\end{figure}

Then the $m$--th classical approximant to the partition function of 
Eq. (\ref{hamilt_iso}) is of the form:
\begin{equation}
\mathcal Z_m=\langle b \mid ({\mathcal W}_1{\mathcal W}_2)^{(N-1)/2} 
\mid a \rangle \quad {\rm for}
\quad {\rm odd} \quad N \, ,
\label{odd}
\end{equation}
\begin{equation}
\mathcal Z_m=\langle b \mid ({\mathcal W}_1{\mathcal W}_2)^{N/2} 
\mid a \rangle  \quad {\rm for}
\quad {\rm even} \quad N  \, ,
\label{even}
\end{equation}
where the operators ${\mathcal W}_1$ and ${\mathcal W}_2$ are defined: 

\begin{equation}
{\mathcal W}_1 = \left( {\mathcal P^2} {\mathcal V_{1,2}}
\right)^{m} \, , 
{\mathcal W}_2 = \left( {\mathcal P^2} {\mathcal V_{2,3}}
\right)^{m} \, 
\label{W}
\end{equation}
and a unitary shift operator:

\begin{equation}
{\mathcal P} \equiv \sum_{S^z_1} \ldots \sum_{S^z_{2m}} \mid 
S^z_2 S^z_3 \ldots S^z_{2m} S^z_1 \rangle \langle S^z_1 S^z_2 
S^z_3 \ldots S^z_{2m} \mid .
\label{P}
\end{equation}
For each finite $N$ the specific heat is given as second derivative of 
the free energy
with respect to temperature, where the free energy is given by the 
formula ${\mathcal F} = - \,k_B T \ln \mathcal Z$.

In order to improve the accuracy of the extrapolations the analysis of the 
specific heat $C_m$ as a function of $1/m^2$ was made using a function 
described by the extrapolation polynomial of the degree $k$
($k=1, \dots , k_{max}$)

\begin{equation}
f(1/m^2)=a_0+ \sum_{j=1}^{k}a_j\cdot\left(\frac{1}{m^2}\right)^j ,
\end{equation}
where the extrapolated value of specific heat $C/T = a_0$.
This extrapolation method was invented in \cite{prb2009rm} for 
the infinite spin chains and will be used in this paper to calculate the
specific heat for the finite spin chains occurring in the case of the 
diluted system (Yb$_{1-x}$Lu$_x$)$_4$As$_3$.

\section{The zero-field specific heat of diluted sample}

To describe the magnetic specific heat of diluted (Yb$_{1-x}$Lu$_x$)$_4$As$_3$ 
we need to calculate the contribution $C_L$ of a finite chain with $L$ sites. 
Assuming the uniform distribution of non--magnetic Lu--ions 
among the chains, each site in the Yb$^{3+}$--chain is randomly 
occupied by a magnetic ion with a probability $p=1-x$. The probability
of two arbitrary sites being occupied is $p^2$. The probability 
of one end having an empty neighbour is $(1-p)$ and the probability of 
finding cluster with $L$ sites is $p^L (1-p)^2$ . The total number 
of $L$-clusters is $n_L=N p^L (1-p)^2$ ($N \rightarrow \infty$ is the total 
chain length and is much larger than the cluster length). The total number of 
all $L$-clusters ($L = 1, \cdots , \infty $) is given by the following sum:
\begin{eqnarray}
n_t = \sum_{L=1}^{\infty} n_L = 
N \sum_{L=1}^{\infty} p^L (1-p)^2 = 
\nonumber \\
=N (1-p)^2 \sum_{L=1}^{\infty} p^L = N (1-p) p.
\end{eqnarray}
Using the probability distribution of chains with $L$ sites
\begin{equation}
w_L = \frac{n_L}{n_t} = \frac{N p^L (1-p)^2}{N p (1-p)} = p^{L-1} (1-p)
\end{equation}

we obtain the specific heat per spin \cite{Asakawa}:

\begin{equation}
C=\frac{n}{N} \sum_{L=1}^{\infty} w_L C_L ,
\end{equation}
where $C_L$ denotes the specific heat of a finite chain with $L$ sites,
$N$ is the number of all the spins in the system and $n$ is the number 
of chains (where $n/N = x = (1-p)$).

We note that the specific heat per mole of 
(Yb$_{1-x}$Lu$_x$)$_4$As$_3$ is given as $(1-x) C$ and finally 
we have

\begin{eqnarray}
C = x \sum_{L=1}^{\infty} p^{L-1} (1-p) (1-x) C_L = 
\nonumber \\
=x^2 \sum_{L=1}^{\infty} (1-x)^L C_L ,
\label{C}
\end{eqnarray}
where $C_L=L C_{sp}(L)$ and $C_{sp}(L)$ denotes specific heat per spin.
According to the formula (\ref{C}) the specific heat of the diluted 
system is determined by all the possible finite size contributions $C_L$ which 
seems unfeasible to achieve. However, for $N$ large enough we enter the asymptotic 
region where $C_L$ varies as $1/L$ which is demonstrated in Fig.~\ref{c_l}.

\begin{figure}
\begin{center}
\includegraphics*[scale=0.44]{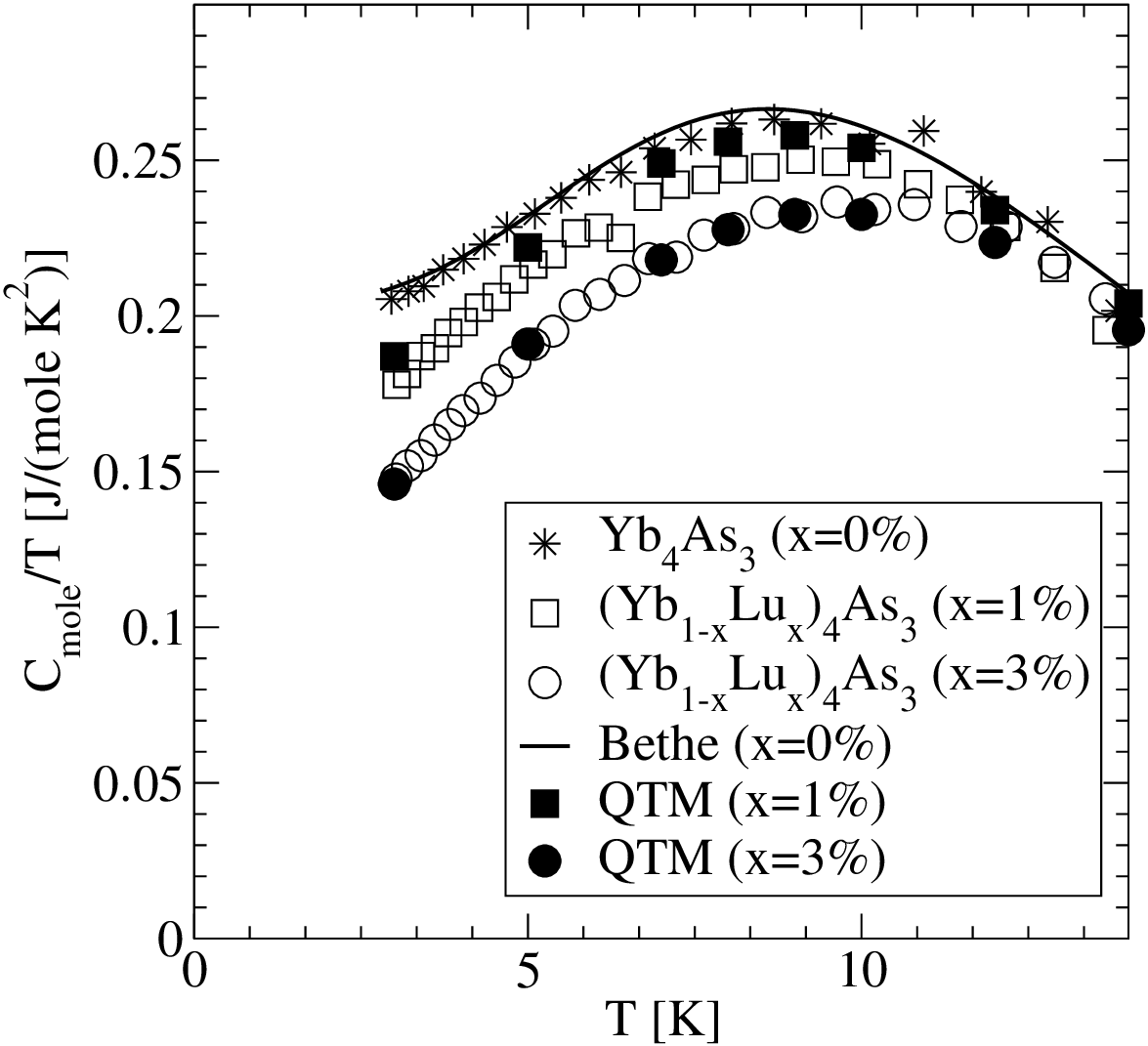}
\end{center}
\caption{Temperature dependence for the zero-field specific heat 
of (Yb$_{1-x}$Lu$_x$)$_4$As$_3$ for different impurity concentrations $x$. 
Open symbols represent experimental data for $x \neq 0$ and the corresponding filled 
symbols indicate the simulation results. The experimental and 
numerical zero-field data for pure Yb$_4$As$_3$ are 
also presented by asterisks and the continuous line, respectively.}
\label{C_T}
\end{figure}

For each temperature we have calculated the specific heat $C_{sp}(L)$ for 
$L \leq 30$. Our specific heat results for two temperatures 
($T=7$ and $T=14$ K) are shown in Fig. \ref{c_l}. The open 
symbols represent the specific heat for various numbers of sites $L$. 
The filled symbols represent zero-field specific heat data $C_{BA}$ 
for infinite chains, which are exact Bethe ansatz results
\cite{Jonston}. For sufficiently large $L > L_0$ we can 
estimate the specific heat by the linear function:

\begin{equation}
C_{sp}(L)=C_{BA}+a \frac{1}{L} ,
\label{C_lin}
\end{equation}
so that

\begin{eqnarray}
C(L>L_0)=x^2 \sum_{L=L_0+1}^{\infty} (1-x)^L 
L \left( C_{BA} + \frac{a}{L} \right) = 
\nonumber \\
= x^2 C_{BA} \sum_{L=L_0+1}^{\infty} (1-x)^L 
 L +x^2 \sum_{L=L_0+1}^{\infty} a (1-x)^L .
\label{C1}
\end{eqnarray}

Finally, using in Eq. (\ref{C1}) the formula for 
the geometric series, we have specific heat for $L>L_0$:

\begin{eqnarray}
C(L>L_0)=C_{BA} (1-x)^{L_0+1} (L_{0} x + 1 ) + 
\nonumber \\
+ a x (1-x)^{L_{0} +1}
\label{C2}
\end{eqnarray}

and specific heat for whole range of $L$:

\begin{eqnarray}
C=x^2 \cdot \sum_{L=1}^{L_0} (1-x)^L \cdot C_{sp}(L) L + C(L>L_0)
 .
\label{C3}
\end{eqnarray}

\begin{figure}

\begin{center}

\includegraphics*[scale=0.38]{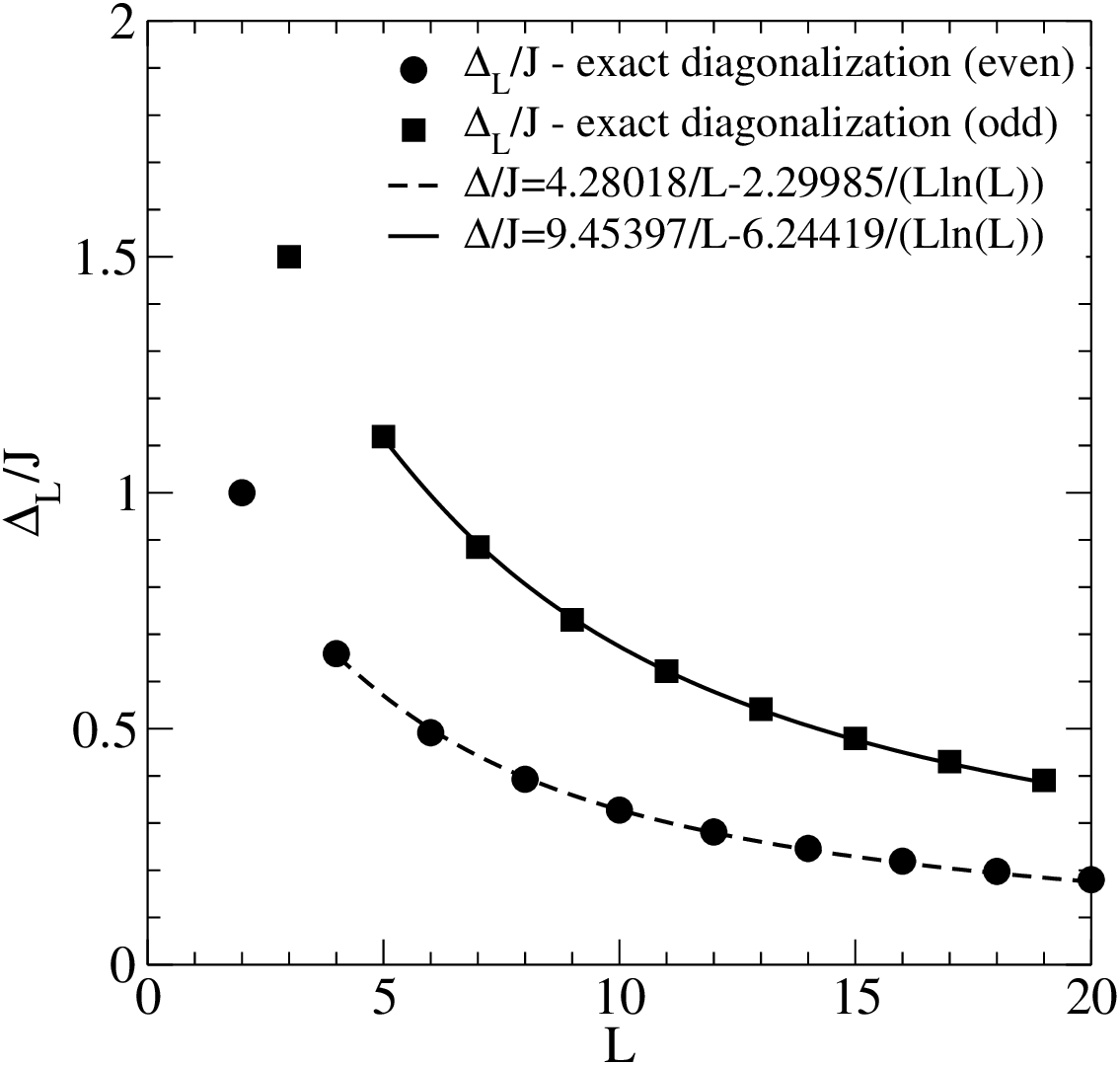}

\includegraphics*[scale=0.38]{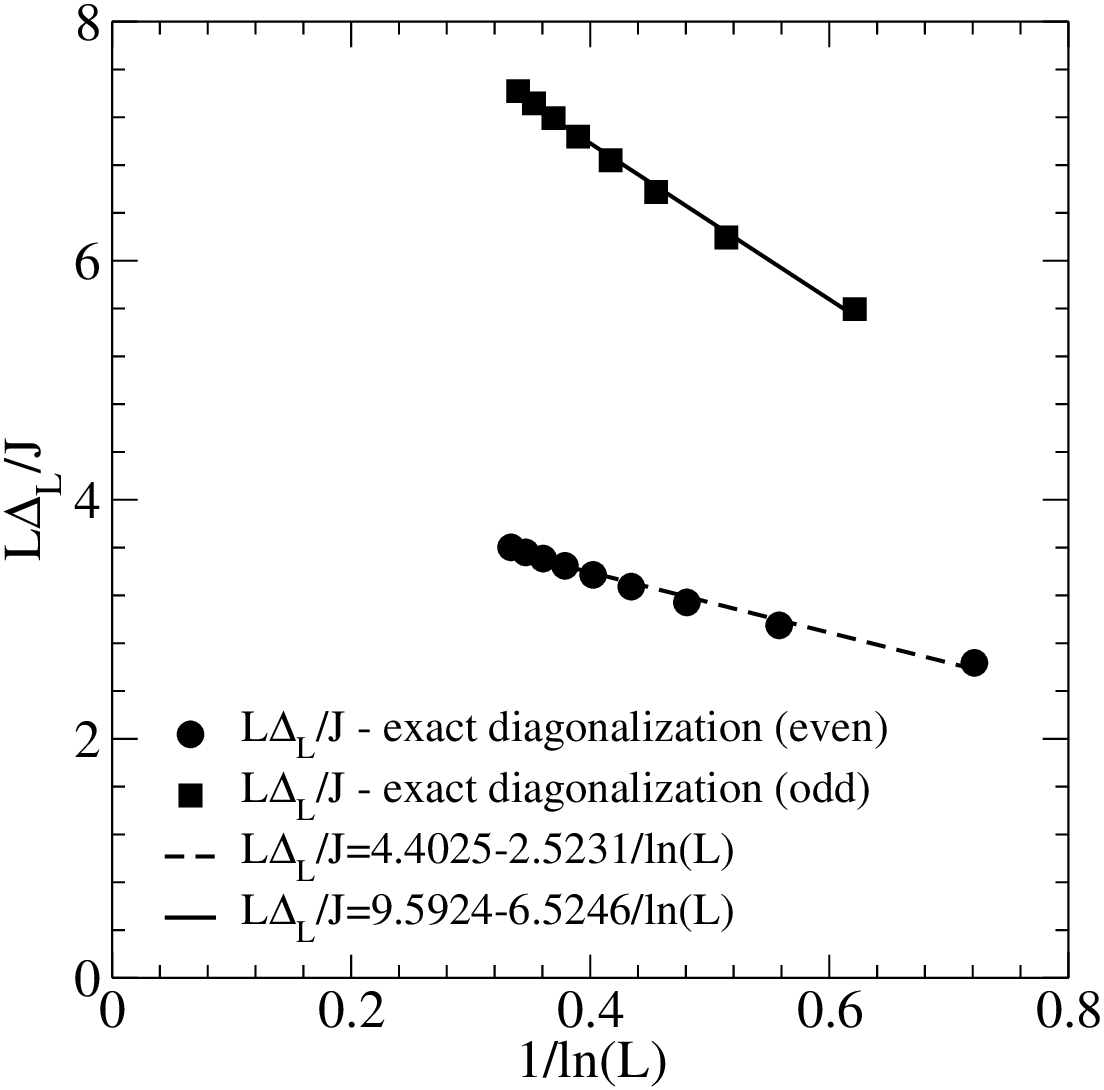}

\includegraphics*[scale=0.38]{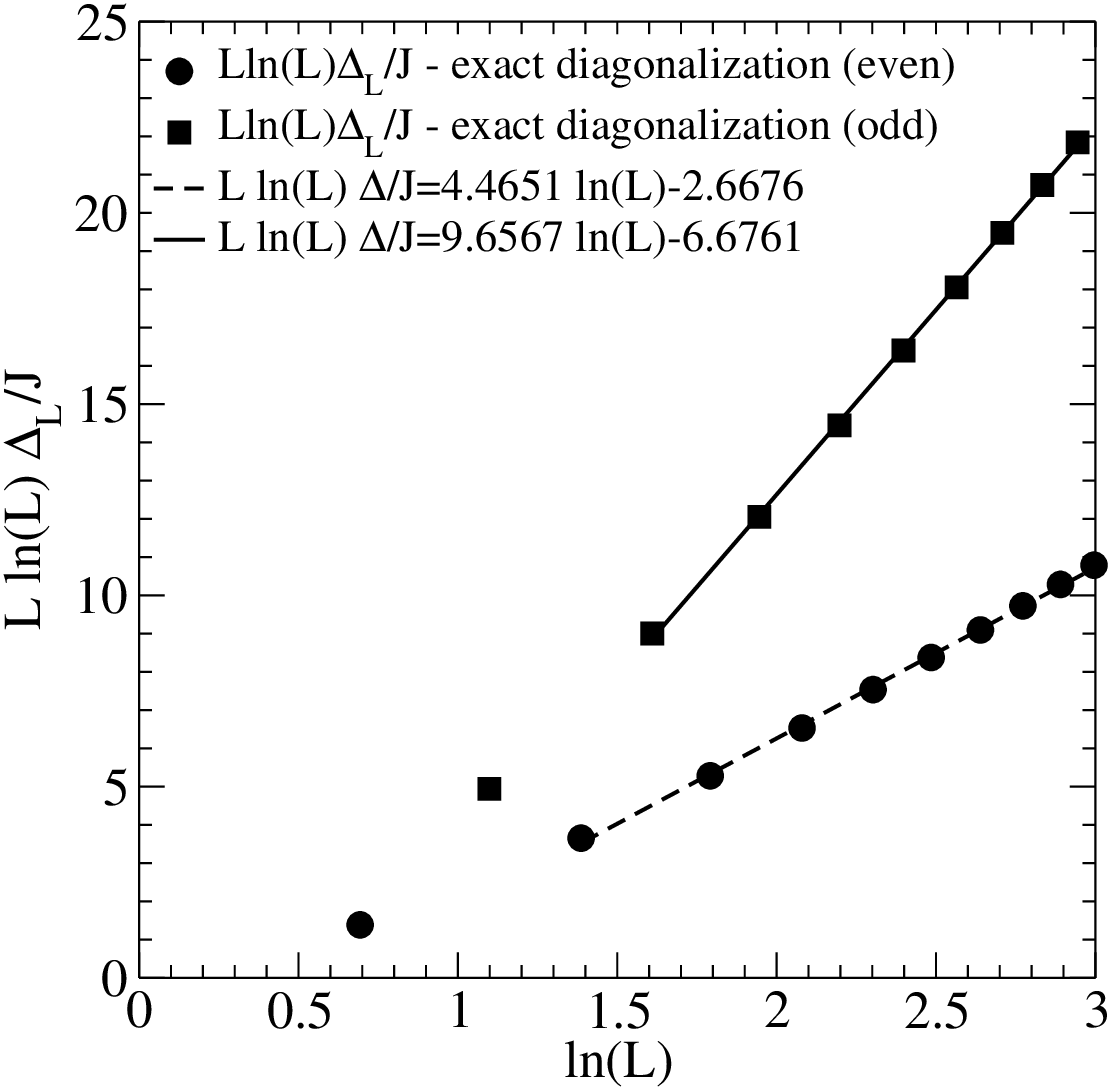}

\end{center}

\caption{Size dependence of the energy gaps ($2 \le L \le 20$) 
for the cases I, II, III. 
The circles represent estimates for even number of sites $L$ and 
the 
line is the result of a fit to the corresponding expression. 
The infinite system is gapless.}
\label{Delta_L}
\end{figure}
 
In Fig. \ref{C_T} we compare experimental data and numerical
results obtained from Eq.~(\ref{C3}) for the zero-field specific 
heat of (Yb$_{1-x}$Lu$_x$)$_4$As$_3$ 
($x=0.01$ and $x=0.03$).  A very good quantitative agreement is 
found in the temperature region analyzed
so that the strong reduction of the heat capacity with $x$ is 
nicely reproduced within our theoretical model.

\section{Average energy gap}

Similar calculations can be performed to obtain the average energy 
gap for a given concentration of impurities. If we define $\Delta_L$ 
as energy gap for the segment with length 
$L$, the average energy gap is described by the following equation:

\begin{equation}
\Delta/J = x^2 \cdot \sum_{L=1}^{\infty} (1-x)^L
 \Delta_L/J .
\label{Delta}
\end{equation}

The energy gaps for ${ L \leq 20}$ were calculated within the exact 
diagonalization technique and are presented as a function of $L$ by the symbols in 
Fig. \ref{Delta_L}a. In the case of even number of spins per segment $L$ the 
ground state is a nonmagnetic singlet, whereas it is a magnetic doublet for 
odd $L$~\cite{Haas}. In both cases $\Delta_L$ represents the gap from the 
respective ground to the first excited states.

As expected, the gaps $\Delta_L$ decrease with the size $L$ and their asymptotic behavior 
\begin{equation}
\Delta_L/J = \frac{\alpha}{L} + \frac{\beta}{L\; \mathrm{ln}(L)},
\label{fit_delta}
\end{equation}
is known from the conformal field theory (CFT) approach
\cite{Haas,Affleck_Qin,Haas_1998}, where  $\alpha = \pi^2/2=4.93$
\cite{Cardy} for the the even $L$. We consider also the modified expressions
\begin{equation}
\frac{L \Delta_L}{J} = \alpha + \frac{\beta}{ \mathrm{ln}(L)},
\label{fit_alpha}
\end{equation}
and
\begin{equation}
 \mathrm{ln}(L)\frac{L \Delta_L}{J} = {\alpha}\,{\mathrm{ln}(L)} + \beta.
\label{fit_beta}
\end{equation}

In the limit of very long segments the values $\alpha,\, \beta$ found from 
Eqs.~(\ref{fit_delta})-(\ref{fit_beta}) should be the same, but for smaller 
$L$ they may be different and also depend on the number of data points considered. 
The expressions~(\ref{fit_delta})-(\ref{fit_beta}) are represented in 
Figs.~\ref{Delta_L}a, \ref{Delta_L}b and \ref{Delta_L}c, respectively, and are 
also referred to as the cases I, II, III. In the fits presented, only the lowest 
values $L=2,3$ were discarded. The corresponding estimates of the parameters 
$\alpha,\, \beta$ are quoted in the legend of a given figure. For the even-numbered 
segments $\alpha =4.28,\, 4.40,\, 4.47$ in the case I, II, III, respectively. 
If the logarithmic correction is neglected ({\it i.e.} $\Delta_L/J = \frac{\alpha}{L}$) 
and two data points with the highest $L$ = 18 and 20 are taken into account 
(the case IV), then $\alpha$ = 4.78, $\beta$ = -3.52. All these estimates of 
$\alpha$ are consistent with the CFT value $\alpha$ = 4.93.

\begin{figure}
\begin{center}
\includegraphics*[scale=0.38]{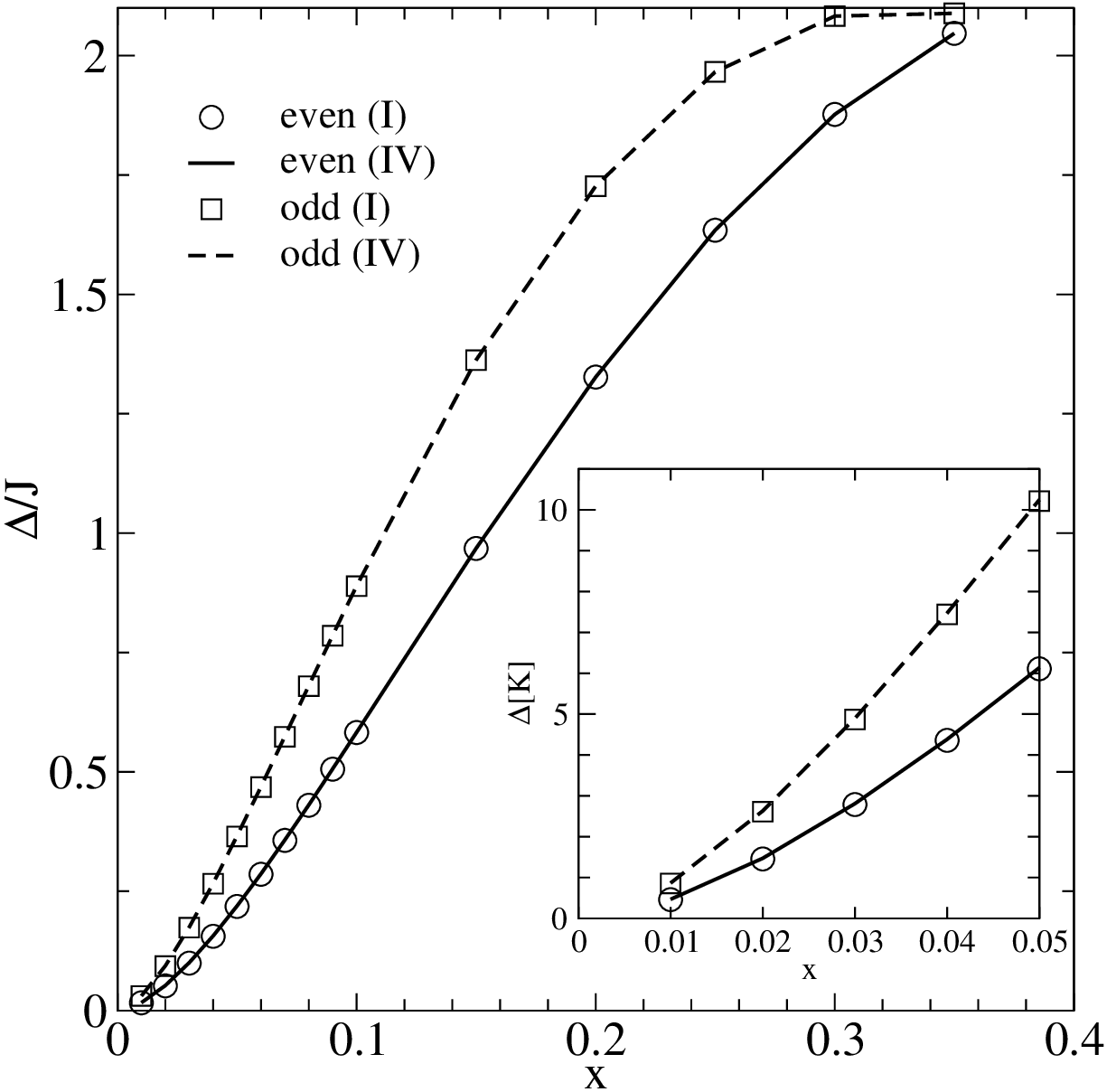}
\end{center}
\caption{Dependence of the excitation energy on the concentration of non-magnetic 
impurities $x$. In the inset the low concentration part of the dependence is 
enlarged and expressed in the absolute units.}
\label{Delta_x}
\end{figure}

Having found the coefficients $\alpha$ and 
$\beta$ determining the asymptotic behavior of $\Delta_L/J$, the dependence of 
the average energy gap on the concentration of impurities $x$ can be estimated 
from Eq.~(\ref{Delta}). 
Excitation energies for the even and odd segments are plotted in  Fig.\ref{Delta_x} 
for the cases I and IV.
For each type of segment, the results are not sensitive to the values $\alpha$ and 
$\beta$, suggesting that the contribution from the long segments is small. 
The average energy gap coincides with that calculated for even segments.
For small concentrations $x \sim 1-3\% $ the average gap is small,
nevertheless its impact on the specific heat is strong and leads to 
its substantial reduction.

\section{Scaling behavior}

It is known from literature and our previous paper \cite{prb2009rm} that  
pure Yb$_4$As$_3$ is an almost ideal example of a linear Heisenberg 
antiferromagnet. Here we have also proven that in the diluted system the 
experimental specific heat data can be described by the same Heisenberg model. 
This implies that our compound with non-magnetic randomly distributed impurities 
is a good realization of segmented Heisenberg spin-1/2 chains.

For these chains the following low-temperature scaling behavior \cite{Haas} 
of the specific heat was found in terms of the scaling variable $(T/J)^{-1/2}$
\begin{equation}
\mathrm{ln} \left( C \left( \frac{T}{J} \right)^{5/4} \right)
= \Phi -\gamma \left( \frac{T}{J} \right)^{-1/2},
\label{eq_gamma}
\end{equation}
where $\gamma$ is the $x$-dependent amplitude and $\Phi$ is a constant independent 
of the scaling variable. The scaling variable depends on the ratio $T/J$ so that 
the dependence (\ref{eq_gamma}) can be verified provided that the magnetic 
coupling constant is known which in our case amounts to $J/k_B$ = 28 K.

\begin{figure}
\begin{center}
\includegraphics*[scale=0.44]{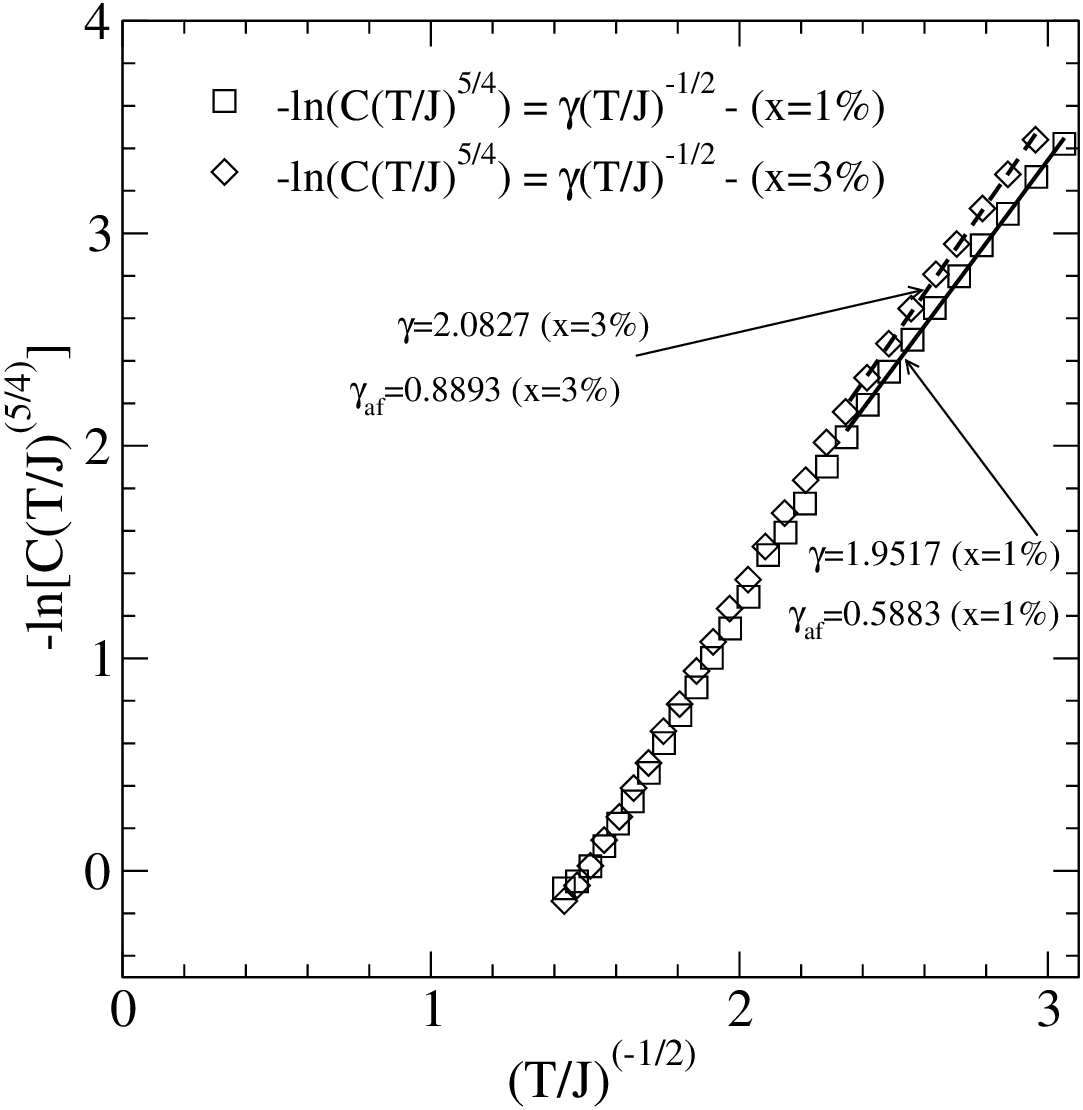}
\end{center}
\caption{The scaling behavior of the low temperature 
specific heat of diluted Yb$_4$As$_3$ for impurity concentrations 
$x$ = 1\% and 3\%. The linear dependence is observed for the highest 
values of the scaling variable.}
\label{fig_gamma}
\end{figure}

In Fig.~\ref{fig_gamma} the experimental results rescaled according to 
Eq.~(\ref{eq_gamma}) are plotted as a function of the scaling variable. 
In the lowest temperature region they display the expected linear dependence. 
This indicates that our experimental results confirm the scaling behavior of the 
low-temperature specific heat. This fact is important on its own but it also 
independently confirms our choice of magnetic coupling constant.

The $\gamma$ values estimated are listed in Fig.~\ref{fig_gamma} and compared 
with $\gamma_{af}$ calculated from the analytical formula quoted in 
literature \cite{Haas}. They agree qualitatively but do not coincide. 
We attribute the deviations to the limited validity of the analytical 
formula which was derived for high concentrations.

We note that the scaling law (\ref{eq_gamma}) is fulfilled only for 
finite $x$. In the limit $x$=0 corresponding to the infinite S-1/2 spin 
chains, the low temperature specific heat is linear in $T$ so that 
Eq.~(\ref{eq_gamma}) cannot be satisfied.

\section{Conclusions}

We have proven that the spin model worked out for the pure compound 
Yb$_4$As$_3$ can also explain the specific heat data in the presence 
of non-magnetic impurities, assuming their random distribution. 
We have applied the quantum transfer matrix method to obtain the accurate 
numerical results without any adjustable parameter.

This conclusion implies that the diluted compound 
(Yb$_{1-x}$Lu$_x$)$_4$As$_3$ studied here is a good realization of 
segmented antiferromagnetic Heisenberg spin-1/2 chains with frozen 
impurities and the low temperature specific heat data can be used to 
verify experimentally the theoretically expected scaling law. We have demonstrated 
that the lowest temperature tail of the logarithm of the rescaled specific 
heat  as a function of $(T/J)^{-1/2}$ (Eq.~(\ref{eq_gamma})) fulfills 
the linear dependence predicted by conformal field theory. 
This finding also provides independent evidence for the correct choice 
of the magnetic coupling constant which is unchanged compared to the pure system. 
We have also confirmed that the amplitude 
of the leading term determining the energy gap as a function of the 
size $L$ for even segments is consistent with the earlier theoretical 
prediction. The impurity driven reduction of the specific heat can be attributed 
to the 
finite energy gaps appearing in the even-numbered segments.

\begin{acknowledgments}
P.G. likes to acknowledge discussions with Andreas Honecker. This work was 
supported by the Polish Ministry of Science and High
Education grant N202 230137 and was granted access to the HPC 
resources in PSNC Pozna\'n (Poland) and those available within 
DECI program by the PRACE-2IP under grant no RI-283493
\end{acknowledgments}

\end{document}